\documentclass[aps,twocolumn,floatfix,floats,showpacs,superscriptaddress,raggedbottom]{revtex4-1}
\usepackage{graphicx,latexsym}
\usepackage{dcolumn}
\usepackage{amsmath}
\usepackage{amssymb,bm}
\usepackage[normalem]{ulem}

\def\mb#1{\mbox{\boldmath$#1$}}

\def\eq#1{Eq.\ (\ref{#1})}
\def\fig#1{Fig.\ \ref{#1}}

\usepackage{hyperref}
\hypersetup{
    pdfnewwindow=true,       
    colorlinks=true,         
    linkcolor=blue,          
    citecolor=blue,          
    filecolor=magenta,       
    urlcolor=black           
}

\usepackage{booktabs}

\begin{document}


\title{Cavity-photon-switched coherent transient transport \\ in a double quantum waveguide}

\author{Nzar Rauf Abdullah}
\email{nra1@hi.is}
\affiliation{Science Institute, University of Iceland,
        Dunhaga 3, IS-107 Reykjavik, Iceland}

\author{Chi-Shung Tang}
 \affiliation{Department of Mechanical Engineering,
        National United University, 1, Lienda, Miaoli 36003, Taiwan}

\author{Andrei Manolescu}
 \affiliation{Reykjavik University, School of Science and Engineering,
              Menntavegur 1, IS-101 Reykjavik, Iceland}

\author{Vidar Gudmundsson}
\email{vidar@raunvis.hi.is}
 \affiliation{Science Institute, University of Iceland,
        Dunhaga 3, IS-107 Reykjavik, Iceland}

%

\begin{abstract}
We study a cavity-photon-switched coherent electron transport in a symmetric
double quantum waveguide. The waveguide system is weakly connected
to two electron reservoirs, but strongly coupled to a single quantized photon cavity
mode. A coupling window is placed between the waveguides to allow for electron
interference or inter-waveguide transport. The transient electron transport in 
the system is investigated using a quantum master equation. 
We present a cavity-photon tunable semiconductor quantum waveguide implementation of
an inverter quantum gate in which the output of the waveguide system may be selected 
via the selection of an appropriate photon number, or 'photon frequency' of the cavity. 
In addition, the importance of the photon polarization in the cavity that is either parallel 
or perpendicular to the direction of electron propagation in the waveguide system is demonstrated.
\end{abstract}



\maketitle


\section{Introduction}

In quantum information technology researchers seek quantum storage devices to develop a quantum computer
in which a qubit is used as an elementary unit for encoding information.
In practice, several systems have been suggested  
to built a qubit. Among many based on semiconductors promising are, for example, 
double quantum dots~\cite{PhysRevLett.107.030506} 
and double quantum waveguides (DQW)~\cite{ApplPhysLett.81.22}.

A semiconductor waveguide can be defined as a quantum wire conserving the phase coherence of electrons
in the system at low temperature~\cite{Ionicioiu.15.125}.
Two parallel quantum waveguides, separated by an electrostatic potential barrier and 
coupled via a coupling region or a window to facilitate an interference between
the waveguides, may be one of the candidates to construct a qubit~\cite{PhysRevLett.84.5912}.
The characteristics of the transport of electrons through the double waveguide system
determines possible quantum logic operations~\cite{SpperlatticeandMicrostructure.2.30}.
A Not-operation is realized if an electron switches from the first waveguide 
to the second waveguide~\cite{NANOTECHNOLOGYIEEE.1.3}, and a square-root-of-Not-operation ($\sqrt{\rm NOT}$)
is formed if the electron wave splits equally between the waveguides~\cite{PhysRevA.70.052330}.

Several proposals have been suggested to control the electron motion in a waveguide system
that provides the qubit operation such as:
Magnetic switching, an external magnetic field can be used to transfer
an electron wave between two asymmetric waveguides~\cite{ApplPhysLett.79.14},  
Electrostatic potential switching, the coupling window can be defined by
a saddle potential that washes out fluctuation resonance peaks and 
increases the speed of electron switching processes between the waveguides~\cite{NANOTECHNOLOGYIEEE.6.5},
A single quantum dot close to the coupling window has been considered to 
enhance electron inter-waveguide transport
in a Coulomb blockade regime~\cite{ApplPhysLett.86.052102},
and Electron switching by using acoustic waves~\cite{SemicondSciTechnol.19.412}.

There are still non-trivial aspects that need to be investigated 
concerning the control of electron switching in a DQW system for implementing an action of a quantum logic gate.
In this work, we show how a cavity-photon can implement a quantum logic gate action
in a single semiconductor qubit that is embedded in a photon cavity with a single quantum mode.
A qubit system can be constructed from a coupled double semiconductor waveguide.
Our DQW system consists of symmetric control- and target-waveguides
with a window is placed between them to facilitate inter-waveguide transport. 
The DQW is weakly connected to two leads with asymmetric coupling where 
the left lead is coupled only to the control-waveguide while the right lead 
is connected to both the control- and the target-waveguide.
The DQW system is in a photon cavity in which
the photons can be polarized parallel ($x$-direction)
or perpendicular ($y$-direction) to the direction of electron propagation
with a fixed electron-photon coupling strength. A non-Markovian quantum master
equation is used to explore the electron transport through the DQW system
caused by a bias between the external leads~\cite{Vidar11.113007,PhysRevB.82.195325}.

In the absence of a photon cavity, we observe oscillations
in the charge current by varying the length of the coupling window (CW).
The oscillations are caused by inter-waveguide transport due to
interference of states between the guides. 
In the presence of the photon cavity, the current oscillations are affected by the 
photon polarization, the number of photons, and the photon frequency
as these cavity parameters influence the interference between the states in the waveguide system. 
Therefore, the electron motion can be manipulated to implement a quantum logic gate action 
by the cavity photon. In the $x$-polarized photon field
a $\sqrt{\rm NOT}$-operation and a NOT-operation quantum logic gates are obtained 
by tuning the photon number in the cavity.
For the $y$-polarized photon field, electron-switching process can be
accomplished if the energy of 'absorbed photon(s)' is equal or greater than
the confinement energy of electron in that direction in the waveguide system.
The many-body cavity-photon-switching is of importance as yet another mechanism to implement
quantum logic gate operations in a semiconductor qubit.

The paper is organized as follows: In Sec.~\ref{Sec:II}, we present
the model describing the window-coupled double waveguide system based on the quantum master equation (QME)
approach. Section \ref{Sec:III} presents our numerical results and
discussion. Concluding remarks are addressed in Sec.~\ref{Sec:IV}.

\section{Model and Theoretical Method}\label{Sec:II}

\begin{figure}[htbq!]
 \includegraphics[width=0.42\textwidth,angle=0]{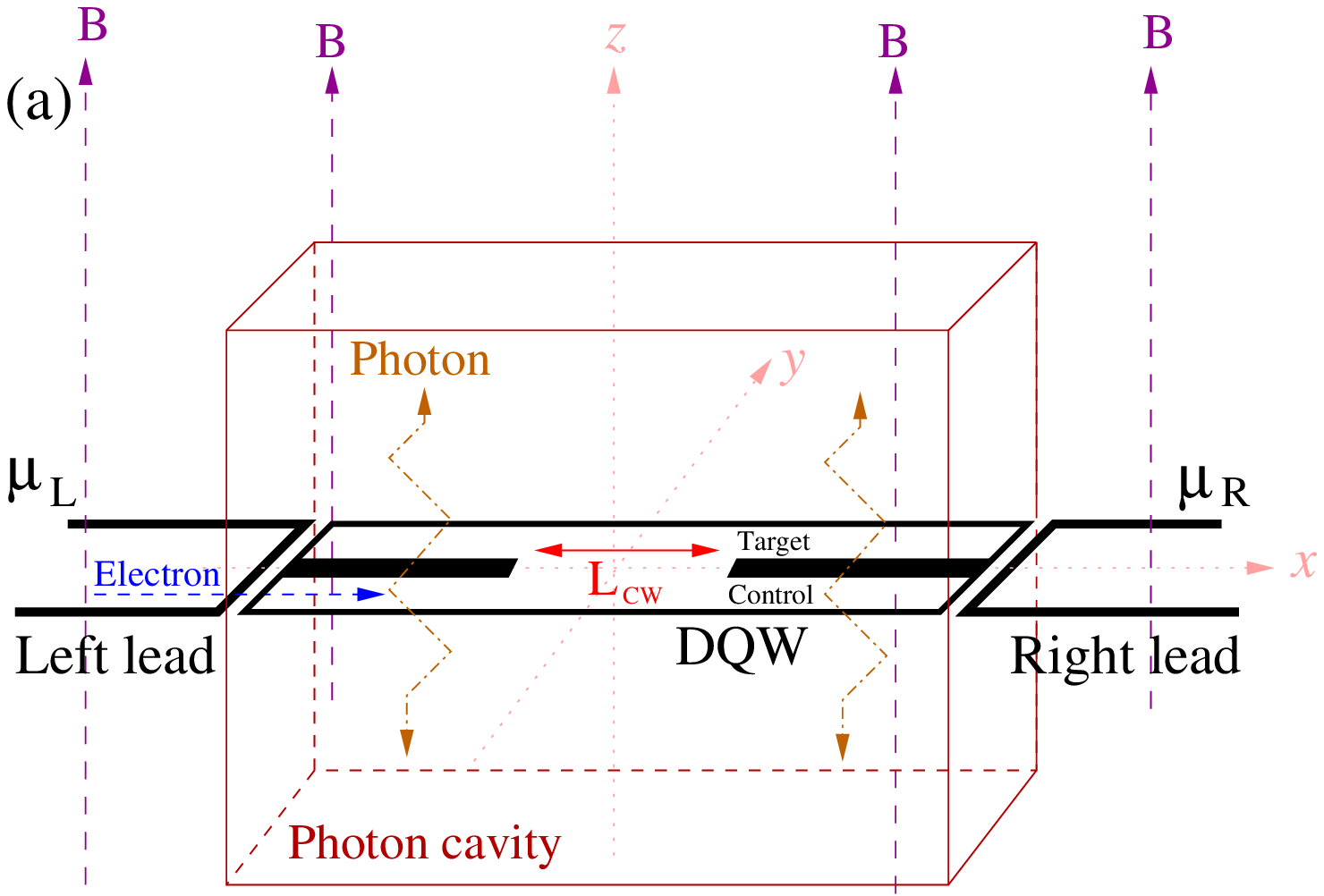}\\
 \includegraphics[width=0.42\textwidth,angle=0]{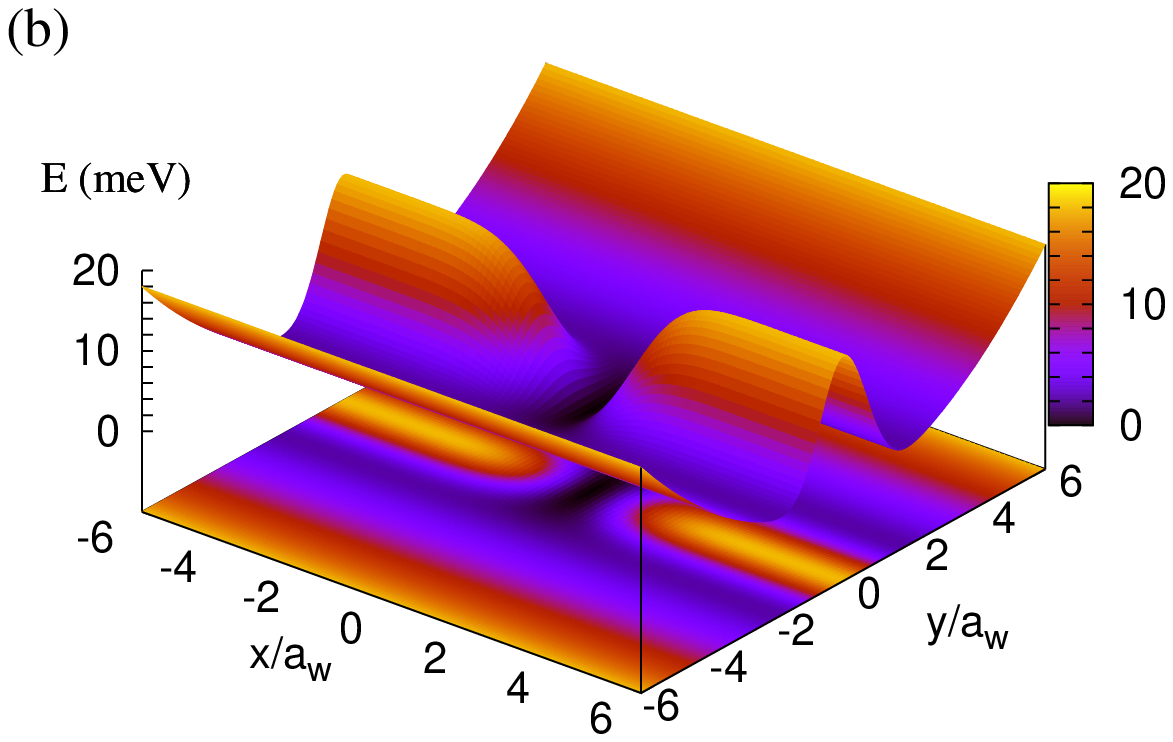}
 \caption{(Color online)
      (a) Schematic of double quantum waveguide (DQW) coupled to a photon cavity, connected to the left lead with chemical
      potential $\mu_L$ and the right lead with chemical potential $\mu_R$ in an external magnetic field B. A coupling window 
      is placed between the control- and the target-waveguide with a length $L_{\rm CW}$ (red arrow).
      The electrons from the left lead only enter the control waveguide (blue arrow).
      (b) The potential defines the double waveguide system with a coupling window between the control- and target-waveguide.  
       All lengths are characterized by the effective magnetic length $a_w$. The physical parameters are
       $B = 0.001$~T, $a_w = 33.72$~nm, $L_{\rm CW}=100$~nm, and $\hbar\Omega_0 = 1.0$~meV.}
      \label{fig01}
\end{figure}

We model a two dimensional symmetric double quantum wire in a perpendicular
magnetic field. 
The double waveguide system is placed in a photon cavity as is schematically shown in \fig{fig01}(a).
The waveguide system is connected to two external leads with different 
chemical potentials $\mu_{l}$ where $l$ refers to the left (L) or
the right (R) lead, respectively. The DQW system consists of a control- and a target-waveguide 
with the same width providing a symmetric double waveguide system. A window is placed
between the waveguides with length $L_{\rm CW}$ (red arrow) to facilitate inter-waveguide transport.
The DQW and the leads are exposed to an external magnetic field $B$ in the $z$-direction.
The total system is designed such that the electrons in the left lead are only injected into  
the control-waveguide (blue dashed arrow). 

Figure \ref{fig01}(b) shows the DQW potential whose dimensions are characterized 
by the effective magnetic length $a_w$.
The DQW system has a hard-wall confinement in $x$-direction at $x= \pm L_x/2$,
where $L_x$ is the length of the waveguide system and parabolic confinement in the $y$-direction
with, $V_{\rm c}(y)=\frac{1}{2} m^*{\Omega^2_\mathrm{0}}y^2$,
where $\hbar\Omega_\mathrm{0}$ is the characteristic energy. 
The DQW potential is described as
\begin{equation}
      V_{\rm DQW}(\mathbf{r}) = V_{\rm B}\; {\rm exp}{(-\beta_0^2y^2)} + V_{\rm CW}\; {\rm exp}{(-\beta_x^2 x^2-\beta_y^2 y^2)}.
      \label{V_DQW}
\end{equation}
The first term of \eq{V_DQW} represents a potential barrier between the quantum waveguide 
with $V_{\rm B}=18.0$~meV and $\beta_{0}=0.03$ nm$^{-1}$. The second term defines the 
potential of the CW with $V_{\rm CW} =-18.0$~meV, and $\beta_y=0.03$~nm$^{-1}$ implying a barrier 
width $W_{\rm B} \simeq 66.5$ nm for the first subband. The CW length 
can be estimated as $L_{\rm CW}=2/\beta_x$ and which influences the electron transport between the waveguides.

\subsection{DQW coupled to Cavity}

Now, we demonstrate how the DQW system is coupled to the photon cavity. The Hamiltonian of the
system can be described by a Many-Body (MB) Hamiltonian that describes the DQW and the photon cavity.
The MB Hamiltonian consists of the electronic DQW including electron-electron interaction,
the photon cavity, and the interaction between the DQW and the photon cavity. 
The Hamiltonian of the total system can be written as

\begin{align}\label{HS}
      H_\textrm{S} &=  \sum_{n,n^{\prime}} \langle\psi_n \lvert \left[ \frac{(\bm{\pi}_e + 
      \frac{e}{c} \mathbf{A}_{\rm ph})^2}{2m^*}
        + V_\mathrm{DQW} \right]\lvert\psi_{n^{\prime}}\rangle  d_n^{\dagger} d_{n^{\prime}}
      \nonumber \\
      & + \frac{1}{2}\sum_{\substack{ nn^{\prime} \\  mm^{\prime}}} (V_{\textrm{Coul}})_{nn',m'm} \
        d_n^{\dagger}d_{n^{\prime}}^{\dagger}d_md_{m^{\prime}} \nonumber \\
      &  + \, \hbar\omega_{\textrm{ph}} a^{\dagger} a + 
      g_{\rm ph}\sum_{n,n^{\prime}}d_n^{\dagger}d_{n^{\prime}}\; g_{nn^{\prime}}
      \left\{a + a^\dagger\right\} 
      \nonumber \\
      & +\frac{g_{\rm ph}^2}{\hbar\Omega_w} \sum_{n}d_n^{\dagger}d_n
      \left[  \hat{N}_{\rm ph} + \frac{1}{2}\left( a^\dagger a^\dagger + aa  + 1 \right)\right].
\end{align}
The first term of \eq{HS} describes the DQW system without the electron-electron interaction,
where $|\psi\rangle$ is a single-electron SE state, $m^*$ is the effective mass of an electron, $e$ is the electron charge
and $d^\dagger_{n}$ and $d_{n^{\prime}}$ are the electron creation and annihilation operators, respectively. 
In addition, $\bm{\pi}_e= p+\frac{e}{c}\mathbf{A}_{\mathrm{ext}}$, where $p$ is the momentum
operator, $\mathbf{A}_{\mathrm{ext}}$ is the vector potential for the static magnetic field 
which can be defined as $\mathbf{A}_{\mathrm{ext}}$ = ($0,-By,0$),
and $\mathbf{A}_{\rm ph}$ is the photon vector potential that can be introduced as 
\begin{equation}
      \mathbf{A_{\rm ph}} = A_{\rm ph}
       \left( a+a^{\dagger} \right)\mathbf{\hat{e}}\, ,
\end{equation}
herein, $A_{\rm ph}$ is the amplitude of the photon field, and $\mathbf{\hat{e}}$
is the unit vector that determines the direction of the photon polarization 
either parallel ($e_x$) in a TE$_{011}$ mode or perpendicular ($e_y$) in a TE$_{101}$ mode
to the transport direction.
In the second term of \eq{HS}, the Coulomb interacting electron Hamiltonian is shown
with the Coulomb matrix elements in the SE state basis $(V_{\textrm{Coul}})_{nn',m'm}$~\cite{PhysRevB.82.195325}.
The third term of \eq{HS} denotes the free photon field, where
$\hbar\omega_{\mathrm{ph}}$ is the quantized photon energy, and $a^{\dagger}(a)$ is
the operator of photon creation (annihilation), respectively.
In \eq{HS}, the para-magnetic ($g_{\rm ph}$-term), and the dia-magnetic ($g^2_{\rm ph}$-term) of
the electron-photon interaction are presented in which $g_{\rm ph} = e A_{\rm ph} \Omega_wa_w/c$ is 
the electron-photon coupling, and $g_{nn^{\prime}}$ are
the dimensionless electron-photon coupling tensor elements~\cite{Vidar85.075306}, and 
$\hat{N}_{\rm ph} = a^{\dagger} a$ is the photon number operator.

In our calculations, we include both the para- and the dia-magnetic interaction terms
which lead to more complex photon-electron interaction processes than are present in the resonant two-level 
Jaynes-Cummings model, where only the paramagnetic term is taken into account~\cite{ProceedingsoftheIEEE.51.89}.
In addition, we use exact diagonalization (configuration interaction) including many levels 
to treat the electron-electron Coulomb interaction and the electron-photon 
interaction~\cite{Yannouleas70.2067, NewJournalOfPhysics.14.013036,Vidar61.305} 
without resorting to the rotating wave approximation~\cite{PhysRevA.70.052315, PhysRevLett.98.013601}.

\subsection{Transport Formalism}

Our model for the calculation of time dependent 
properties of transport through an open system requires a coupling  
to electron reservoirs. Here, we show how the central system is connected 
to the leads via coupling regions. Later in this section, 
a time-dependent formalism will be presented to investigate the electron transport in the system.

The total Hamiltonian of the system describing the waveguide system, the leads, 
and the coupling between the DQW and the leads can be written as
\begin{align}
      H(t) &= H_\textrm{S} + \sum_{l=\mathrm{L,R}}  \int d{\mb{q}}\, \epsilon^l(\mb{q}) {c^l_{\mb{q}}}^\dagger
              c^l_{\mb{q}} \\
           &+ \sum_{l=\mathrm{L,R}}  \chi^l(t) \sum_{n}\int d{\mb{q}}\, \left[  {c^l_{\mb{q}}}^\dagger T^l_{\mb{q}n} d_n
            + d^\dagger_n (T^l_{n\mb{q}})^* c^l_{\mb{q}}\right], \nonumber
\label{Ht}
\end{align}
where $H_\textrm{S}$ indicates the Hamiltonian of the DQW system coupled to the photon cavity shown in \eq{HS}. 
The second term of the Hamiltonian describes the $l^{\mathrm{th}}$ lead with 
$\mb{q}$ being the dummy index representing the momentum of the standing electron waves in the semi-infinite
leads and their subband number~\cite{PhysRevB.82.195325}, $\epsilon^l(\mb{q})$ being the
single-electron energy spectrum in the lead $l$, and 
${c^l_{\mb{q}}}^\dagger$ ($c^l_{\mb{q}}$) being
the electron creation (annihilation) operators, respectively.
The last term of the Hamiltonian demonstrates the time-dependent coupling between the DQW
and the leads describing a transfer of an electron between a single-electron state
of the central system $|n\rangle$ and a single-electron energy state of
the leads $|\mb{q}\rangle$ through a coupling tensor
\begin{equation}
      T^l_{\mb{q}n} =
      \int d\mathbf{r} d\mathbf{r^{\prime}} \psi^l_{\mb{q}}(\mathbf{r}')^*
      g^l_{\mb{q}n} (\mathbf{r},{\bf r'}) \psi^\mathrm{S}_n({\bf r}),
\label{Tlqn}
\end{equation}
with $\psi^\mathrm{S}_n({\bf r})$  ($\psi^l_{\mb{q}}(\mathbf{r}')$) being a single-electron wave
functions of the DQW system (leads). In addition, $\chi^l(t)$ is a time-dependent function
defining the onset of the coupling, and
\begin{eqnarray}
      g^l_{\mb{q}n} ({\bf r},{\bf r'}) &=&
                   g_0^l\exp{\left[-\delta_x^l(x-x')^2-\delta_y^l(y-y'-\alpha)^2\right]}
                   \nonumber \\
                   && \times \exp{\left( -\Delta_{n}^l(\mb{q}) / \Delta
                   \right)} \label{cf}
\end{eqnarray}
is a nonlocal coupling where $g_0$ is the coupling strength, 
$\delta_x^l$ and $\delta_y^l$ are the coupling parameters that 
control the range of the coupling in the $x$- and $y$-direction, respectively,
$\Delta_{n}^l(\mb{q}) = |E_n-\epsilon^l(\mb{q})|$ and $\Delta$ adjust the energy 
overlap of lead and DQW  states and wavefunctions in the 
contact region~\cite{Vidar11.113007}, and $\alpha$ is a skewing parameter 
that shifts the weight of the coupling from the left lead to  
the control-waveguide.

We use a non-Markovian QME formalism to calculate the electron transport from the left lead to the right lead
through the DQW system~\cite{Breuer2002}. The QME approach describing the time-dependent electron transport
can be obtained from quantum Liouville-von Neumann equation~\cite{RevModPhys.81.1665}
\begin{equation}
 \dot{\rho}(t)= -\frac{i}{\hbar}\left[H(t),\rho(t)\right],
\end{equation}
where $\rho(t)$ is the density operator of the total system. 
The total density operator before the coupling between 
the waveguide system and the leads can be written as
$\rho(t_0)$ = $\rho_\mathrm{L}\rho_\mathrm{R}\rho_\mathrm{S}(t_0)$, where
$\rho_\mathrm{L}$ and $\rho_\mathrm{R}$ are the density
operators of the isolated left and right leads, respectively~\cite{Nzar.25.465302}.

Our aim in this work is to seek the dynamics of the electron and 
the inter-waveguide switching processes in the system. 
To calculate the electron motion in the DQW system under the influence of the leads,
we take the trace over the Fock space with respect
to the lead variables to build a reduced density operator of the waveguide system 
$\rho_\mathrm{S}(t)={\rm Tr}_\mathrm{L} {\rm Tr}_\mathrm{R} \rho(t)$,
which leads to the Nakajima-Zwanzig equation of time-evolution in
an open system~\cite{Haake1973}

\begin{equation}
      \dot{\rho_\mathrm{S}}(t) = -i{\cal L}_{\rm S}\rho(t) +\int_{t_0}^{t}
      dt' {\cal K}(t,t')\rho_\mathrm{S} (t'),
\label{NZ-eq}
\end{equation}
where
${\cal L}_{\rm S}\cdot =[H_S,\cdot]/\hbar$ is the Liouvillian with respect to the
time-independent Hamiltonian $H_S$ of the DQW system and
${\cal K}(t,t')$ is the integral kernel representing the dissipative
time-dependent coupling to the leads~\cite{Haake1973,PhysRevB.82.195325}.
For the regime of weak coupling by sequential tunneling to the leads 
treated in our model we derive the dissipative kernel of Eq.\ \ref{NZ-eq} 
by keeping terms up to second order in the time dependent 
coupling~\cite{NewJournalOfPhysics.11.073019,Vidar85.075306}.

The reduced density operator allows us to calculate the left and the right charge currents
into or out of the DQW~\cite{Nzar.25.465302}. Therefore, the net charge current can be introduced as 
\begin{equation}
      I_{\rm Q}(t) = I_{\rm L}(t) - I_{\rm R}(t),
\label{I_Q}
\end{equation}
where $I_{\rm L}(t)$ denotes the partial current from the left lead into
the control-waveguide and $I_{\rm R}(t)$ describes to the partial current
into the right lead from both waveguides~\cite{Nzar.25.465302}.

To explore the properties of the charge switching between
the waveguides, the expectation value of the charge current density operator in the central system 
is calculated. The charge current density can be defined as
\begin{equation}
      \mathbf{J}({\bf r},t) = {\rm Tr} \left( \hat{\rho}_\mathrm{S}(t)  \mathbf{\hat{J}}({\bf r}) \right),
\end{equation}
where the charge current density operator is
\begin{align} \nonumber
  \mathbf{\hat{J}}({\bf r}) &=
         \sum_{nn'} \Bigg( \frac{e\hbar}{2 m^* i} \Big[ \psi^\mathrm{S*}_{n}({\bf r}) (\nabla \psi^\mathrm{S}_{n'}({\bf r}))
         - (\nabla \psi^\mathrm{S*}_{n}({\bf r})) \psi^\mathrm{S}_{n'}({\bf r})  \Big] \\
         &+   \frac{e^2}{m^*} \Big[ \mathbf{A}_{\rm ext}({\bf r}) + \mathbf{A}_{\rm ph}({\bf r}) \Big]  \psi^\mathrm{S*}_{n}({\bf r})
         \psi^\mathrm{S}_{n'}({\bf r}) \Bigg) d_n^{\dagger} d_{n'}.
\end{align}

In the following, we shall investigate numerically the influence of the cavity photon
on the coherent electron transport through the DQW system
in the case of $x$- or $y$-polarization of the photon field.

\section{Results and Discussion}\label{Sec:III}

In this section, we will discuss our numerical results that demonstrate photon-switched
coherent electron transport in a double quantum waveguide. 
To provide coherent electron transport in the system, we consider 
the double waveguide system to be made of a GaAs semiconductor with length $L_x=300$~nm.
It is known that the phase coherence length $L_{\phi}$ of a GaAs semiconductor can be $\sim (30 -
40)\times10^{3}$~nm at low temperature $T \sim 0.1 - 2.0$~K~\cite{Datta1995}.
Thus the coherence length is much larger than the length of the waveguide system
which is an essential requirement to construct a qubit in quantum information technology.

We have fixed the following physical parameters in the calculations, the temperature of the leads is $0.5$~K, 
the chemical potentials of the leads are consider to be $\mu_L = 4.0$~meV and $\mu_R = 3.0$~meV,
the confinement energy of the leads and the DQW system in the $y$-direction is 
$\hbar \Omega_l = 1.0$~meV and $\hbar \Omega_0 = 1.0$~meV, respectively,
the skewing parameter is $\alpha = 4 a_w$, and the electron-photon coupling strength is $g_{\rm ph} = 0.1$~meV.

\subsection{The system without/with photon cavity}

In order to understand the influence of the photons on the transport 
we first explore the electron transport characteristics in the system without and with the photon cavity.
Initially, the photon energy and the electron-photon coupling strength are assumed to be constant 
at $\hbar \omega_{\rm ph} = 0.3$~meV and  $g_{\rm ph} = 0.1$~meV, respectively.

In a previous work~\cite{arXiv:1408.1007}, we demonstrated the effects of the electron-electron interaction
and an external magnetic field on the electron switching process between the waveguides.
In this work, we will show how photons in a cavity can be used to switch the electron motion between the
waveguides.

\begin{figure}[htbq!]
 \includegraphics[width=0.30\textwidth,angle=0,bb=54 50 211 294,clip]{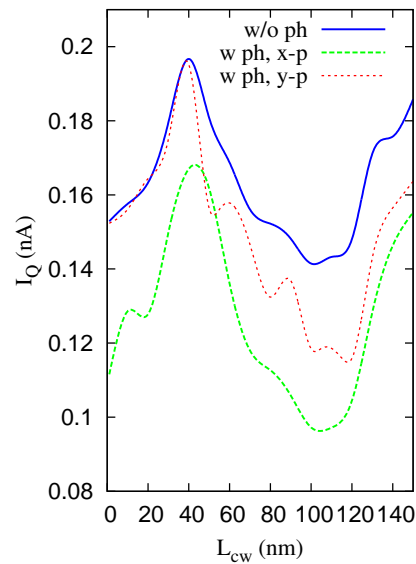}
 \caption{(Color online)
          The net charge current $I_{\rm Q}$ versus coupling window length $L_{\rm CW}$
          without (w/o) a photon (ph) cavity (blue solid), and with a photon (w ph) cavity 
          in the case of $x$-polarization (green dashed) and $y$-polarization (red dotted) at time $t = 200$~ps. 
          The photon energy $\hbar \omega_{\rm ph} = 0.3$~meV,  $g_{\rm ph} = 0.1$~meV, $B = 0.001$~T,
          and the chemical potentials are $\mu_L = 4.0$~meV and $\mu_R = 3.0$~meV, 
          implying $\Delta \mu = 1.0~{\rm meV}$.}
      \label{fig02}
\end{figure}

Figure \ref{fig02} shows the net charge current versus the CW length $L_{\rm CW}$ 
without (w/o) a photon cavity (ph) (blue solid), and with (w) a photon cavity in $x$-polarized ($x$-p) 
(green dashed) and $y$-polarized ($y$-p) (red dotted) photon field.
The oscillation in the net charge current depends on the transport properties of electrons between 
the control- and the target-waveguide. The electrons 
can be subjected to inter-waveguide forward or backward scattering, consequently 
a current peak at $L_{\rm CW}\simeq40$ and a current dip
at $L_{\rm CW}\simeq110$~nm are formed. The net charge current decreases in the presence
of cavity photon for the $x$- and $y$-polarized field 
where the cavity contains one photon initially.

\begin{figure}[htbq!]
  \includegraphics[width=0.23\textwidth,angle=0,bb=54 50 211 294,clip]{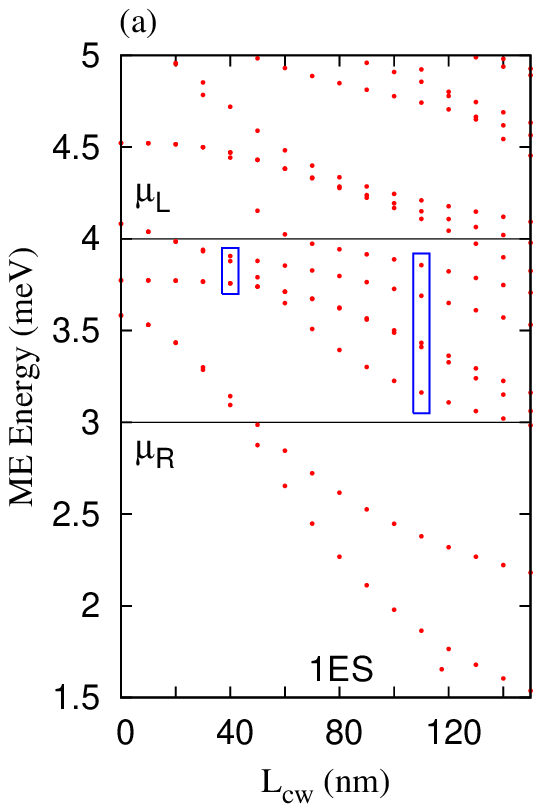}
  \includegraphics[width=0.23\textwidth,angle=0,bb=54 50 211 294,clip]{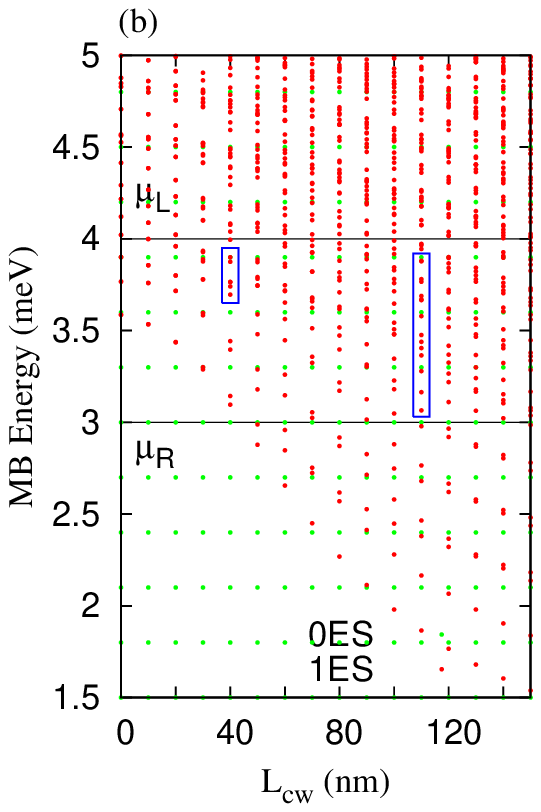}
 \caption{(Color online) Energy spectra  of the DQW system as a function of
          CW length $L_{\rm CW}$ without (a) and with (b) photon cavity in the system
          including zero-electron states (0ES, green dots) and
          one-electron states (1ES, red dots) at $B = 0.001$~T. The 1ES states in the left blue rectangle
          are close to the crossover region of states, but the 1ES states in the right blue rectangular are not.
          The left rectangle contains the most active transport states. The change in the height
          of the rectangle from left to right indicates the spreading of states from the resonance to 
          the off-resonant condition.
          The photon energy $\hbar \omega_{\rm ph} = 0.3$~meV with $x$-polarization, $g_{\rm ph} = 0.1$~meV.
          The chemical potentials are $\mu_L = 4.0\ {\rm meV}$ and $\mu_R = 3.0\ {\rm meV}$ (black)
          implying $\Delta \mu = 1.0~{\rm meV}$.}
\label{fig03}
\end{figure}

To explain the current oscillation and the suppression in the net charge current 
in the presence of photon cavity, we refer to the energy spectrum of the DQW system.
Figure \ref{fig03} shows energy spectra for the DQW system as a function of 
the CW length $L_{\rm CW}$ for the case of the no photon cavity (a), and for 
the system in the photon cavity (b).                   
When the CW length $L_{\rm CW}$ is increased, we observe following effects in the energy spectra:
the energy of the states with an electron component decreases monotonically, 
and generally the degeneration of energy levels reduces.
We observe an energy level crossover at $L_{\rm CW}\simeq40$~nm, and increased splitting of levels at $\simeq110$~nm. 
The weak tunneling through the central barrier between the waveguides leads to almost
degenerate symmetric and antisymmetric one-electron states, but the opening of the coupling window
increases the ``interaction'' between these states leading to a reduced degeneracy.  

In \fig{fig03}(a), the low end of the spectrum with only one-electron states (1ES) (red dots) is shown
for the waveguide system without the photon cavity.
At $L_{\rm CW}\simeq40$~nm (left blue rectangle) higher excited states
enter the active bias window resulting in a level crossover with lower excited states~\cite{PhysRevB.76.195301,ChinPhysLett.24.8}.
The energy crossover reflects a 'resonance' energy levels between the waveguides leading to 
inter-waveguide transport.
The contribution of the higher excited states to the electron transport increases the net charge current
forming a current peak as is shown in \fig{fig02} (blue line).
In the current peak, the charge is transferred from the input to the output of the control-waveguide 
with a slight inter-waveguide forward scattering (not shown)~\cite{arXiv:1408.1007}.
For the regime of increased level splitting at $L_{\rm CW}\simeq110$~nm (right blue rectangle) 
the state of the second subband with lowest energy is the highest state in the blue 
rectangle enters the bias window. Even though the energy splitting indicates an 'off-resonance' between
the waveguides, the mixing of a state from the second subband with the first subband
in the electron transport leads to a stronger coupling between the waveguides.
Here, the charge from the control-waveguide partially switches to the target-waveguide due to inter-waveguide backward
and forward scattering and is partially transferred to the output of the control-waveguide 
(not shown)~\cite{arXiv:1408.1007}.
The inter-waveguide backward scattering decreases the net charge current forming a current dip as is shown
in \fig{fig02} (blue line).
We should note that the two-electron states (2ES) of the energy spectrum are not active 
in the presence of the Coulomb interaction because the electron-electron interaction raises the 2ES well above 
the bias window, consequently the 2ES are effectively blocked~\cite{arXiv:1408.1007}.

Figure \ref{fig03}(b) presents the MB energy spectrum including 
zero-electron states (0ES) (green dots) and 1ES (red dots) 
in the presence of a cavity including one photon initially
with the photon energy $\hbar\omega_{\rm ph} = 0.3$~meV and $x$-polarization. 
The one-electron states of the energy spectrum 
decrease monotonically with increasing CW length while 
the zero-electron states (0ES) remain unchanged.
We can clearly see that photon replicas for electron state appear 
with different photon content. The energy difference between two photon replicas
is close to multiples of the photon energy in case of weak electron-photon coupling~\cite{PhysicaE.64.254}.
The photon replicas of the energy levels at $L_{\rm CW}\simeq40$~nm (left blue rectangle)
and $110$~nm (right blue rectangle) become active in the presence of a photon cavity.
To a lesser extent photon replicas of states originally below the bias region that end up in the 
active bias window also contribute. 
Therefore, more states participate in the electron transport. 
In addition, the shape of the active states 
(and the photon replicas) is influenced by the photon field. The photon field
stretches or polarizes the wavefunctions. 

\begin{figure*}[htbq]
       \begin{center}
       \includegraphics[width=0.34\textwidth,angle=0,bb=89 59 285 219]{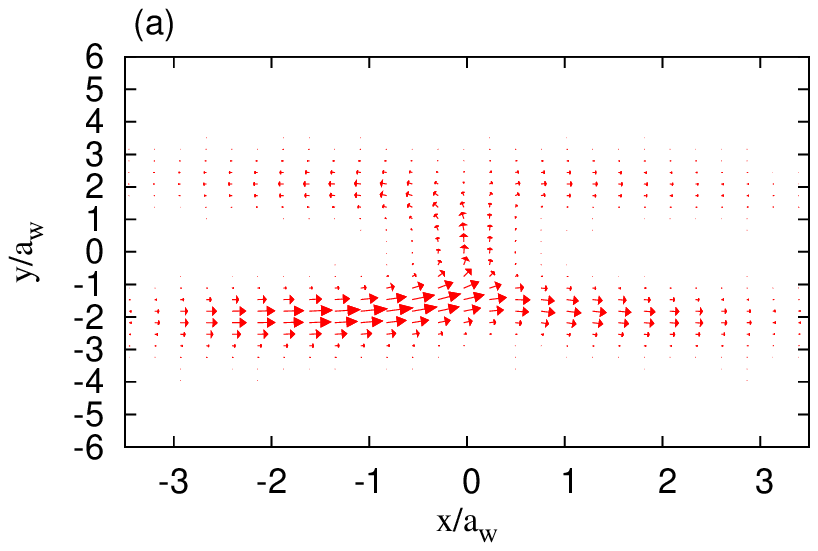}
       \includegraphics[width=0.34\textwidth,angle=0,bb=56 59 252 219]{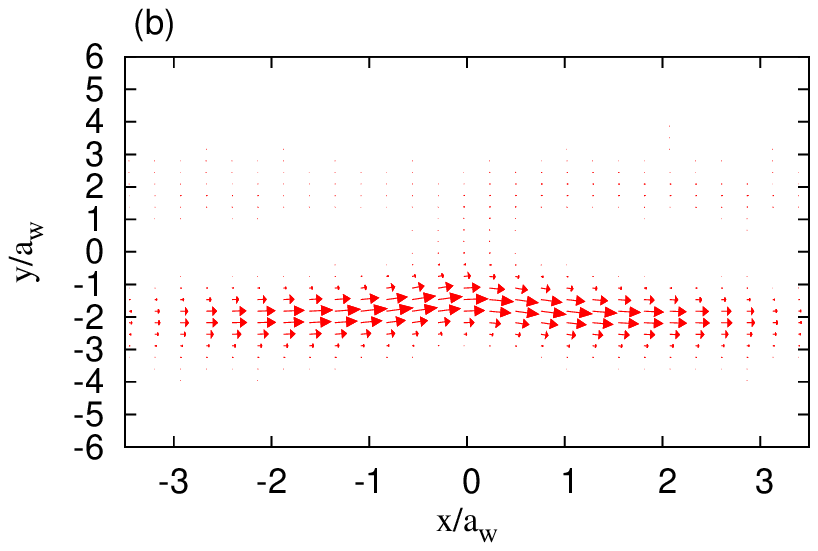}\\  
       \end{center}
       \caption{(Color online) The charge current density at $t=200$~ps 
       for $x$-polarized (a) and the $y$-polarized (b) photon field in the current peak at $L_{\rm CW}\simeq40$~nm  
       shown in \fig{fig02}. The photon energy  $\hbar\omega_{\rm ph} = 0.3$~meV, $N_{\rm ph} = 1$, 
       and  $g_{\rm ph} = 0.1$~meV. The length of the DQW system is $L_x = 300$~nm,
       $\hbar \Omega_0 = 1.0$~meV, $B = 0.001$~T, and $a_w = 33.72$~nm.}
       \label{fig04}
\end{figure*}

The photon replicas have a very important and influential role in the electron-switching process
in the waveguide system. 
At $L_{\rm CW}\simeq40$~nm, the photon replicas of the ground and the first-excited states containing
two photons enter the energy crossover region. The replicas containing 
two photons have a weaker contribution than the replicas containing one photon
in the electron transport because the cavity initially contains only one photon. 
At $L_{\rm CW}\simeq110$~nm (right blue rectangle), a photon replica
of the lowest state of the second subband containing one photon participates in the electron transport. 
The photon replica is a localized state in the CW region leading to
a suppression in the net charge current for both $x$- (green dashed) and $y$-polarized (red dotted) photon field
at the dip as is shown in \fig{fig02}.
However, the photon replicas of the ground state and the first
excited state containing three and four photons are found among the split energy levels.
But they do not influence the electron transport in any significant way.

We should mention that the energy spectrum for the $y$-polarized photon field is very similar
to the spectrum shown in \fig{fig03}(b) with a slightly different photon content in the MB energy states.

Figure \ref{fig04} shows charge current density for the current peak in the $x$-polarized (a), and the $y$-polarized (b)
photon field shown in \fig{fig02} where the photon energy is $\hbar\omega_{\rm ph} = 0.3$~meV 
and the electron-photon coupling is $g_{\rm ph} = 0.1$~meV.
In \fig{fig04}(a), the charge is partially transported through the control-waveguide 
and partially is subject to inter-waveguide backward scattering,
while in the absence of the photon cavity the charge from the input of control-waveguide moves
to the output of the control- and target-waveguide.
The inter-waveguide backward scattering is partially caused by the charge polarization in the $x$-direction
induced by the photon field, and a weak participation of photon replica states containing 
two photons in the electron transport.
As a result, the net charge current decreases in the dip.
In \fig{fig04}(b) the charge remains completely within the control-waveguide because
the photon energy is much smaller than the electron confinement energy in the waveguide system in the $y$-direction.  
The confinement and the photon energy are $\hbar\Omega_0 = 1.0$~meV and 
$\hbar\omega_{\rm ph} = 0.3$~meV, respectively.
In this case, the charge from the control-waveguide does not tunnel into the target-waveguide.
The dynamic evolution of the charge in the control-waveguide implements a controlled NOT function,
which is so called CNOT-operation quantum logic gate leading to enhancement in the net charge current.

\subsection{Variation of the frequency and the initial number of photons}

In this section, we demonstrate how the photon frequency influences
the electron transport through the DQW system in the $x$- and $y$-polarized photon cavity.
In addition, we show the effects of the number of photons initially
in the cavity on electron switching processes between the waveguides.
The electron-photon coupling strength is assumed to be constant at $g_{\rm ph} = 0.1$~meV.

\begin{figure}[htbq!]
\centering
     \includegraphics[width=0.23\textwidth,angle=0,bb=54 50 211 310,clip]{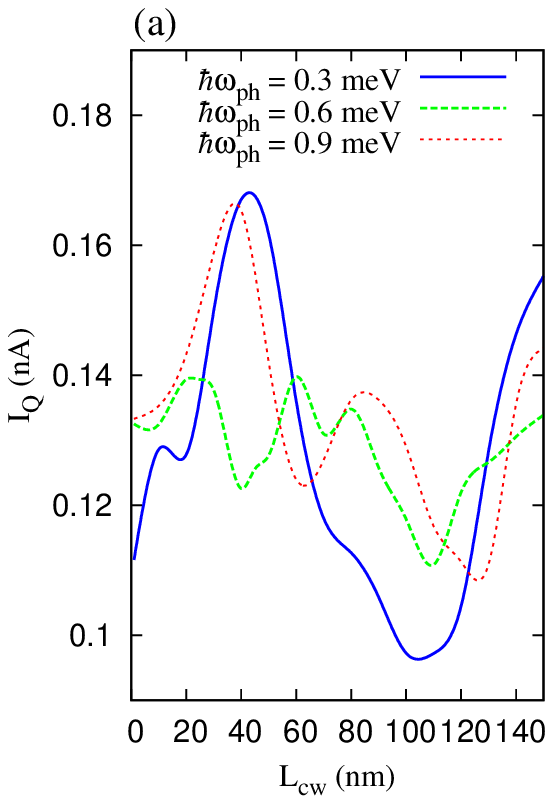}
     \includegraphics[width=0.23\textwidth,angle=0,bb=54 50 211 310,clip]{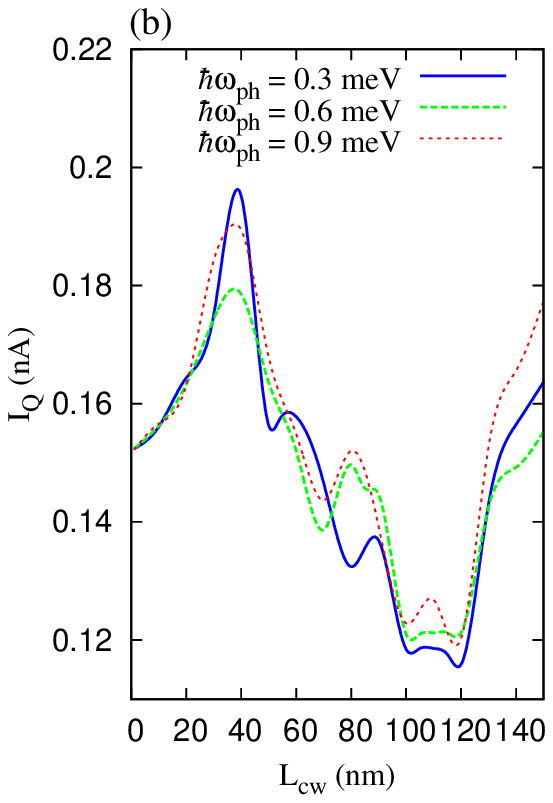}
 \caption{(Color online)
          The net charge current $I_{\rm Q}$ versus coupling window length $L_{\rm CW}$ 
          at time $t = 200$~ps for the $x$-polarized (a) and $y$-polarized (b) photon 
          field with initially one photon in the cavity 
          for different photon energies $\hbar\omega_{\rm ph} = 0.3$~meV (blue solid),
          $0.6$~meV (dashed green) and $0.9$~meV (dotted red).
          The electron-photon coupling $g_{\rm ph} = 0.1$~meV, $B = 0.001$~T,
          and the chemical potentials are $\mu_L = 4.0$~meV and $\mu_R = 3.0$~meV, implying $\Delta \mu = 1.0~{\rm meV}$.}
      \label{fig05}
\end{figure}

Figure \ref{fig05} displays the net charge current for the $x$-polarized (a)
and $y$-polarized (b) photon field with initially one photon in the cavity 
for different photon energies $\hbar\omega_{\rm ph} = 0.3$~meV (blue solid),
$0.6$~meV (dashed green) and $0.9$~meV (dotted red).
In the previous section we discussed the electron transport in the system 
when the photon energy is $\hbar\omega_{\rm ph} = 0.3$~meV for both $x$- and $y$-polarized photon field.
Now, we explore the results when the photon energy is either $\hbar\omega_{\rm ph} = 0.6$~meV or $0.9$~meV. 

We begin by analyzing the net charge current in the $x$-polarized photon field shown in \ref{fig05}(a).
In the case of a photon energy $\hbar\omega_{\rm ph} = 0.6$~meV (dashed green),
the net charge current is strongly reduced for the crossover energy
at $L_{\rm CW}\simeq40$~nm to a current dip instead of the current peak seen for $\hbar\omega_{\rm ph} = 0.3$~meV,
while for the region of split levels at $L_{\rm CW}\simeq110$~nm the net charge current in the dip is enhanced.
If we further increase the photon energy to $0.9$~meV (red dotted), a current peak is again seen 
at $L_{\rm CW}\simeq40$~nm
and a slightly shifted current dip at $L_{\rm CW}\simeq120$~nm.

\begin{figure}[htbq!]
     \includegraphics[width=0.23\textwidth,angle=0,bb=54 50 211 310,clip]{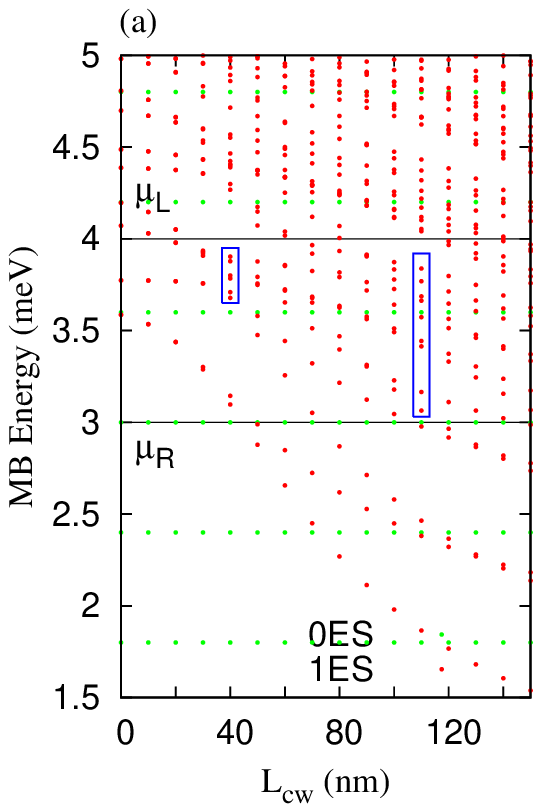}
     \includegraphics[width=0.23\textwidth,angle=0,bb=54 50 211 310,clip]{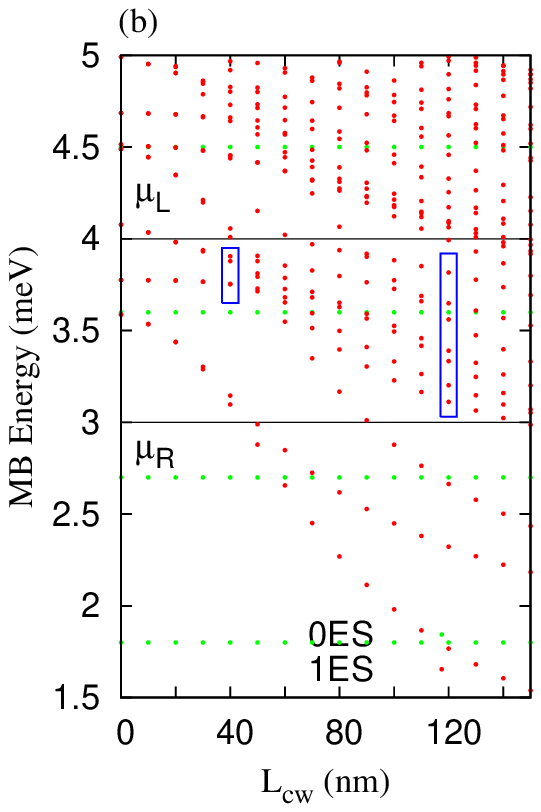}
 \caption{(Color online) Energy spectrum  of the DQW in a cavity as a function of
          CW length $L_{\rm CW}$ and photon energy $\hbar\omega_{\rm ph} = 0.6$ meV (a), and 
          photon energy $\hbar\omega_{\rm ph} = 0.9$ meV (b). The spectra 
          include zero-electron states (0ES, green dots) and
          one-electron states (1ES, red dots) at $B = 0.001$~T. The 1ES in the left blue rectangle
          are close to be in crossing, but the 1ES states in the right blue rectangular are not.
          The left rectangle contains the most active transport states. The change in the height
          of the rectangle from left to right indicates the spreading of states from the resonance to 
          the off-resonant condition.
          The chemical potentials are $\mu_L = 4.0\ {\rm meV}$ and $\mu_R = 3.0\ {\rm meV}$ (black)
          implying $\Delta \mu = 1.0~{\rm meV}$.}
\label{fig06}
\end{figure}

To explore the characteristics of the net charge current in the $x$-polarized photon field, 
we provide \fig{fig06} which shows the MB energy spectrum including zero-electron states (0ES) (green dots) 
and 1ES (red dots) with the photon energy $\hbar\omega_{\rm ph} = 0.6$~meV (a)
and $0.9$~meV (b). 
In \fig{fig06}(a) the MB energy spectrum is shown for the photon energy $\hbar\omega_{\rm ph} = 0.6$~meV.
Each MB state has photon replica with
a different photon content in the presence of the photon cavity.
We notice that the MB ground state is replicated into the CW with one photon 
and energetically enters the region of levels crossover at $L_{\rm CW}\simeq40$~nm (left blue rectangle).
The effect of this localized photon replicated state here on the electron transport is a suppression of the 
net charge current leading to a current dip.
But at $L_{\rm CW}\simeq110$~nm (right blue rectangle), the photon replica of the first excited state
containing one photon contributes to the electron transport leading to an increasing net charge current at the dip 
shown in \fig{fig05} (green dashed).
In \fig{fig06}(b) the MB energy spectrum is displayed for the photon energy $\hbar\omega_{\rm ph} = 0.9$~meV.
The photon replica of neither the ground state nor the first excited states enter the  
active bias window (left blue rectangle).
The result is that the net charge current is almost unaltered.
However, at $L_{\rm CW}\simeq120$~nm (right blue rectangle) the photon replica of the first excited state is found
among the active split energy levels.
This photon replica containing one photon enhances the net charge current in the dip.

\begin{figure}[htbq]
       \begin{center}
       \includegraphics[width=0.34\textwidth,angle=0,bb=89 59 285 219]{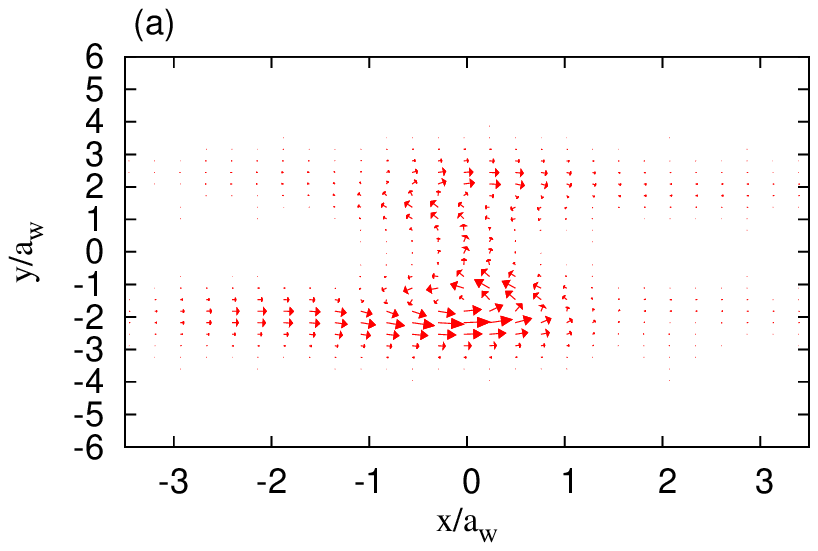}\\
       \includegraphics[width=0.34\textwidth,angle=0,bb=89 59 285 219]{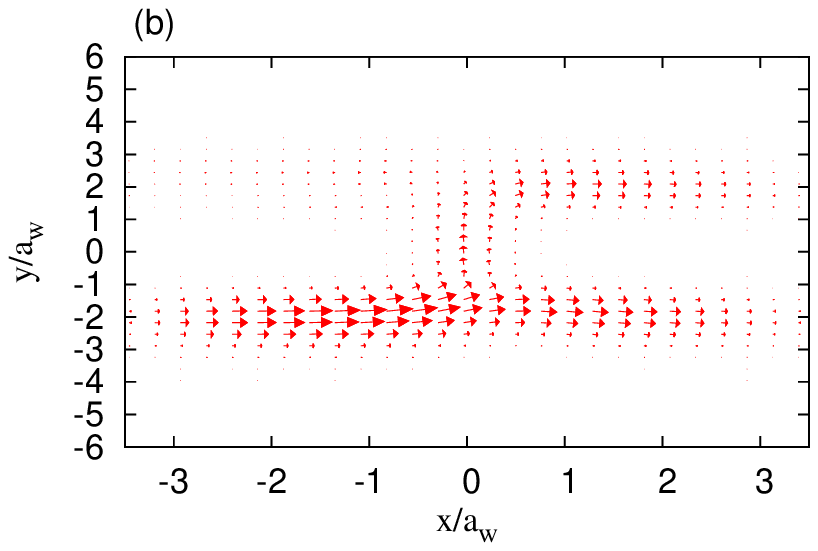} 
       \end{center}
       \caption{(Color online) Charge current density at $t=200$~ps 
                for $x$-polarized photon field with
                photon energy $\hbar\omega_{\rm ph} = 0.6$~meV (a)
                and $0.9$~meV (b) for the current peak at $L_{\rm CW}\simeq40$~nm
                shown in \fig{fig05}. 
                The initial photon number in the cavity $N_{\rm ph} = 1$,
                and the electron-photon coupling $g_{\rm ph} = 0.1$~meV.
                The length of the DQW system is $L_x = 300$~nm, $\hbar \Omega_0 = 1.0$~meV, 
                $B = 0.001$~T, and $a_w = 33.72$~nm.}
       \label{fig07}
\end{figure}

To clarify further the dynamic motion of the charge and explain the current oscillations,
we present \fig{fig07} which shows the charge current density at the current peak shown in \fig{fig05}
in the case of the photon energy $\hbar\omega_{\rm ph} = 0.6$~meV
(a), and $\hbar\omega_{\rm ph} = 0.9$~meV (b). 
In \fig{fig07}(a) the charge current density is seen for the current dip at $L_{\rm CW}\simeq40$~nm when the 
photon energy is $\hbar\omega_{\rm ph} = 0.6$~meV. The charge is localized in the CW region which suppresses
the net charge current and leads to a current dip. The localized charge can be
identified as a contribution of the photon replica of the MB ground state containing one photon.

In \fig{fig07}(b) the charge current density for the current peak at $L_{\rm CW}\simeq40$~nm is presented for
photon energy $\hbar\omega_{\rm ph} = 0.9$~meV. 
The charge from the input control-waveguide is equally split
between the output of the control- and the target-waveguide. 
The photon replica of neither the ground state nor the first excited state 
contribute to the electron transport. But the charge density of the active states
occupies both the control- and target waveguide. Therefore, the net charge current remains almost unchanged.
The splitting of the charge indicates a $\sqrt{\rm NOT}$-operation quantum logic gate action.

We have seen that the charge current density for the current dip at $L_{\rm CW}\simeq110$~nm in the case of 
photon energy $\hbar\omega_{\rm ph} = 0.6$~meV and $0.9$~meV is delocalized (not shown) while a localized charge 
is observed for $\hbar\omega_{\rm ph} = 0.3$~meV. The delocalization of charge is due to
participation of a photon replica of the first excited MB state. Consequently, the net charge current is enhanced.

\begin{figure}[htbq]
       \includegraphics[width=0.45\textwidth,angle=0]{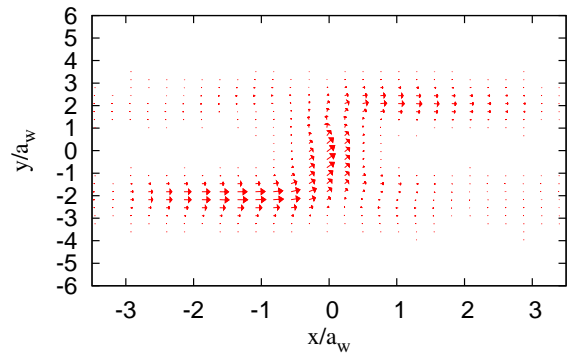}
       \caption{(Color online) Charge current density at $t=200$~ps with $x$-polarized photon field at $L_{\rm CW}\simeq40$~nm.
        The photon energy $\hbar\omega_{\rm ph} = 0.6$~meV, $g_{\rm ph} = 0.1$~meV and $N_{\rm ph} = 2$.  
        The length of the DQW system is $L_x = 300$~nm, $\hbar \Omega_0 = 1.0$~meV, $B = 0.001$~T, and $a_w = 33.72$~nm.}
       \label{fig08}
\end{figure}

We have noticed that the electron-switching process can be achieved by tuning the photon number initially in the cavity.
Let us consider two photons initially in the cavity with energy $\hbar\omega_{\rm ph} = 0.6$~meV and
photon-electron coupling strength $g_{\rm} = 0.1$~meV. Figure \ref{fig08} shows the charge current density 
at the CW length $L_{\rm CW}\simeq40$~nm in the presence two photons in the cavity.
Comparing to the charge current density in the case of one photon in the cavity
with photon energy $\hbar\omega_{\rm ph} = 0.6$~meV shown in \fig{fig07}(a),
the charge motion in the DQW system is drastically changed.
The electron charge switches totally from the input control- to the output target-waveguide.
The dynamic evolution occurring in the DQW system implements a quantum logic gate operation.
In this case, a Not-operation is realized by transferring the charge from the control- to the target-waveguide. 
The electron switching process is due to contribution of a photon replica of the 
both MB ground state and first-excited state containing two photons to the transport.

\begin{figure}[htbq!]
      \includegraphics[width=0.45\textwidth,angle=0]{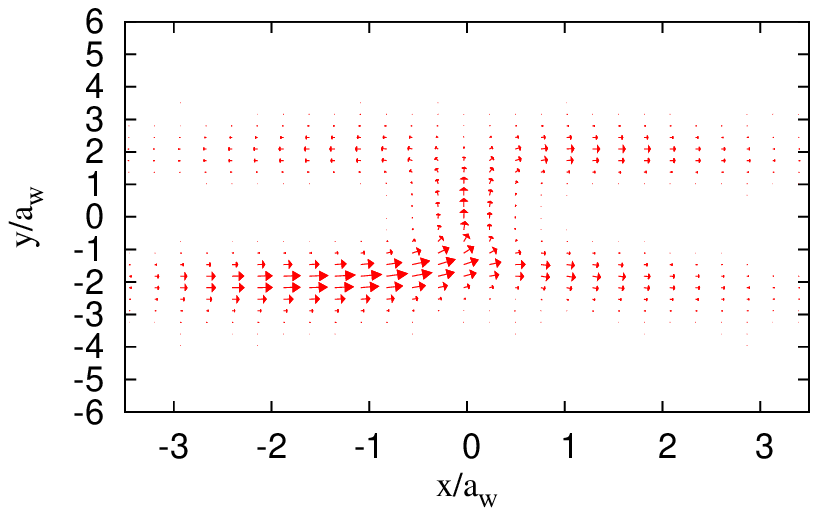}
 \caption{(Color online)
        Charge current density at $t=200$~ps with $y$-polarized photon field in the current peak at $L_{\rm CW}\simeq40$~nm
        shown in \fig{fig05}(b) (red dotted). The photon energy
        $\hbar\omega_{\rm ph} = 0.9$~meV, $g_{\rm ph} = 0.1$~meV and $N_{\rm ph} = 1$.  
        The length of the DQW system is $L_x = 300$~nm, $\hbar \Omega_0 = 1.0$~meV, $B = 0.001$~T, and $a_w = 33.72$~nm.}
      \label{fig09}
\end{figure}

Let's now look at the influences of photon frequency in the $y$-polarized photon field 
on the electron-switching process.
The net charge current $I_{\rm Q}$ in the presence of $y$-polarized photon field 
and initially one photon in the cavity displayed in \fig{fig05}(b)
indicates that the influences of photon frequency on the electron transport is very weak 
compared to the $x$-polarized photon field for the same selected photon energies 
$\hbar \omega_{\rm ph} = 0.3$~meV (blue solid), $0.6$~meV (green dashed) and $0.9$~meV (red dotted).
This is due to the anisotropy of the geometry of the DQW system. 
The total charge-switching from the control- to the target-waveguide can not be achieved
for the same selected photon energy as in the case of a $x$-polarized photon field. For example, \fig{fig09} shows
the charge current density in the current peak at $L_{\rm CW}\simeq40$~nm demonstrated in \fig{fig05}(b)(red dotted), 
where the cavity initially contains one photon and the
photon energy is $\hbar \omega_{\rm ph} = 0.9$~meV.
Comparing to the charge current density shown in
\fig{fig04}(b) when the photon energy is $\hbar \omega_{\rm ph} = 0.3$~meV, inter-waveguide transport is enhanced
because the photon energy here is $\hbar \omega_{\rm ph} = 0.9$~meV, 
which is close to the electron confinement energy in the waveguide system
in the $y$-direction. An electron in the control-waveguide may obtain energy from the photon to partially 
occupy a state in the second subband of the two parallel waveguides and thus being transferred 
to the target-waveguide.

\begin{figure}[htbq!]
      \includegraphics[width=0.45\textwidth,angle=0]{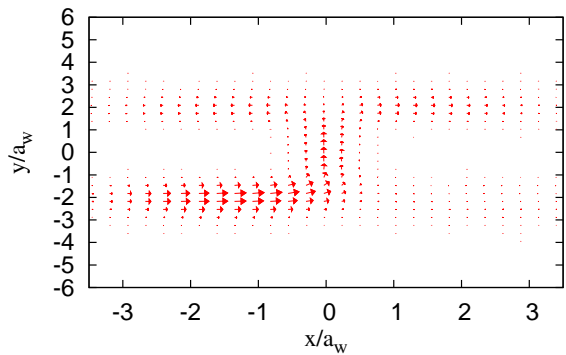}
 \caption{(Color online)
        Charge current density at $t=200$~ps with $y$-polarized photon field 
        at $L_{\rm CW}\simeq40$~nm.
        The photon energy
        $\hbar\omega_{\rm ph} = 0.6$~meV, $N_{\rm ph} = 2$, and $g_{\rm ph} = 0.1$~meV.  
        The length of the DQW system is $L_x = 300$~nm, $\hbar \Omega_0 = 1.0$~meV, $B = 0.001$~T, and $a_w = 33.72$~nm.}
      \label{fig10}
\end{figure}

In order to facilitate total electron-switching between the waveguides
in the $y$-polarization, we need to increase either the photon energy to 
be equal to or greater than the confinement energy of the electrons in the waveguide system
in the $y$-direction or the photon number initially present in the cavity.
We now consider the photon energy to be $\hbar \omega_{\rm ph} = 0.6$~meV, which is
smaller than the electron confinement energy, 
($\hbar\Omega_0 = 1.0$~meV) and consider two photons ($N_{\rm ph} = 2$) initially in the cavity.
An electron in the control-waveguide can absorb two photons with total energy 
$N_{\rm ph}\times\hbar \omega_{\rm ph}\simeq1.2$~meV
and then being transferred to the target-waveguide.
In this case, the charge from the input control-waveguide totally switches to the target-waveguide 
as is shown in \fig{fig10}.
As a result a NOT-operation quantum logic gate action is obtained.

Our results for the two different photon polarizations
have revealed that a variety of quantum logic gate actions can be observed 
in the waveguide system with the switching being strongly 
influenced by the photon energy and the photon number initially in the cavity.

\section{Conclusions and Remarks}\label{Sec:IV}

We have presented the results of a detailed investigation of how to implementing a quantum
logic gate action in a semiconductor qubit by using a new and different technique, 
a cavity-photon-switching. 
In the cavity-photon-switching method, a quantized photon cavity can 
be used to realize a different quantum logic gate actions by varying the photon number, the photon energy, 
or the photon polarization. 

To built a semiconductor qubit, we have considered two parallel symmetric quantum waveguides, 
the control- and the target-waveguide. A window is placed between them to facilitate interference and
inter-waveguide electron transport. The waveguide system is connected to two leads with asymmetric 
coupling in which the control-waveguide is coupled to the leads from both ends while the target-waveguide
is only coupled to the right lead. The DQW system is embedded in a quantized photon cavity with 
a photon field polarized either parallel or perpendicular to the direction of electron motion in 
the system in which the electron-photon interactions is described by exact numerical diagonalization.
We use a non-Markovian master master equation to investigate the transient electron motion in the system.

In the absence of a photon cavity, the electron-switching process depends on 
the ME states active in the electron transport and their
characteristics.
By tuning the CW length, the energy spectrum of the DQW system monotonically decreases
and new states enter and leave the bias window.
Therefore, oscillation in the net charge current occur and indicating inter-waveguide 
forward or backward scattering into the target waveguide. 

In the presence of the photon cavity, photon replicas for each MB energy state
appear. The character of the active photon replicas in the electron transport depend on the photon energy,
the photon number, and the photon polarization in the cavity
In the case of an $x$-polarized photon field, photon replicas contribute to the electron transport
processes leading to the following scenarios:
First,  at high photon energy and one photon initially in the cavity a $\sqrt{\rm NOT}$-operation 
quantum logic gate is found which is due to lifting the photon replica of the ground state out of
the active energy states in the electron transport.
Second, the charge from input the control-waveguide switches to the output of the target-waveguide
in the presence of two photons in the cavity. In this case, an electron in the control-waveguide may 
interact with two photons and transfer to the target-waveguide. 
Therefore, a Not-operation quantum logic gate is implemented.
For the $y$-polarized photon field, the electron-switching processes 
only occur if the photon energy
is equal to or greater than the electron confinement energy in the DQW system in the $y$-direction.

We have demonstrated that the transport properties of a system with nontrivial geometry can be
strongly influenced by choosing the type of electron states replicated into the active transport
bias window. This control can both be excised with the photon energy and the number of photons
in the cavity at the onset of an operation. It should also be stressed that our study of the time-evolution
of the switching and charging processes shows that it is not necessary to await the steady state in order
to complete an operation.

\begin{acknowledgments}
This work was financially supported by the Icelandic
Research and Instruments Funds, the Research Fund
of the University of Iceland, the Nordic High Performance
Computing facility in Iceland, and the Ministry
of Science and Technology, Taiwan through Contract No.\
MOST 103-2112-M-239-001-MY3.
\end{acknowledgments}

%

\bibliographystyle{apsrev4-1}

\begin{thebibliography}{30}
 \makeatletter
\providecommand \@ifxundefined [1]{%
 \@ifx{#1\undefined}
}%
\providecommand \@ifnum [1]{%
 \ifnum #1\expandafter \@firstoftwo
 \else \expandafter \@secondoftwo
 \fi
}%
\providecommand \@ifx [1]{%
 \ifx #1\expandafter \@firstoftwo
 \else \expandafter \@secondoftwo
 \fi
}%
\providecommand \natexlab [1]{#1}%
\providecommand \enquote  [1]{``#1''}%
\providecommand \bibnamefont  [1]{#1}%
\providecommand \bibfnamefont [1]{#1}%
\providecommand \citenamefont [1]{#1}%
\providecommand \href@noop [0]{\@secondoftwo}%
\providecommand \href [0]{\begingroup \@sanitize@url \@href}%
\providecommand \@href[1]{\@@startlink{#1}\@@href}%
\providecommand \@@href[1]{\endgroup#1\@@endlink}%
\providecommand \@sanitize@url [0]{\catcode `\\12\catcode `\$12\catcode
  `\&12\catcode `\#12\catcode `\^12\catcode `\_12\catcode `\%12\relax}%
\providecommand \@@startlink[1]{}%
\providecommand \@@endlink[0]{}%
\providecommand \url  [0]{\begingroup\@sanitize@url \@url }%
\providecommand \@url [1]{\endgroup\@href {#1}{\urlprefix }}%
\providecommand \urlprefix  [0]{URL }%
\providecommand \Eprint [0]{\href }%
\providecommand \doibase [0]{http://dx.doi.org/}%
\providecommand \selectlanguage [0]{\@gobble}%
\providecommand \bibinfo  [0]{\@secondoftwo}%
\providecommand \bibfield  [0]{\@secondoftwo}%
\providecommand \translation [1]{[#1]}%
\providecommand \BibitemOpen [0]{}%
\providecommand \bibitemStop [0]{}%
\providecommand \bibitemNoStop [0]{.\EOS\space}%
\providecommand \EOS [0]{\spacefactor3000\relax}%
\providecommand \BibitemShut  [1]{\csname bibitem#1\endcsname}%
\let\auto@bib@innerbib\@empty
\bibitem [{\citenamefont {van Weperen}\ \emph {et~al.}(2011)\citenamefont {van
  Weperen}, \citenamefont {Armstrong}, \citenamefont {Laird}, \citenamefont
  {Medford}, \citenamefont {Marcus}, \citenamefont {Hanson},\ and\
  \citenamefont {Gossard}}]{PhysRevLett.107.030506}%
  \BibitemOpen
  \bibfield  {author} {\bibinfo {author} {\bibfnamefont {I.}~\bibnamefont {van
  Weperen}}, \bibinfo {author} {\bibfnamefont {B.~D.}\ \bibnamefont
  {Armstrong}}, \bibinfo {author} {\bibfnamefont {E.~A.}\ \bibnamefont
  {Laird}}, \bibinfo {author} {\bibfnamefont {J.}~\bibnamefont {Medford}},
  \bibinfo {author} {\bibfnamefont {C.~M.}\ \bibnamefont {Marcus}}, \bibinfo
  {author} {\bibfnamefont {M.~P.}\ \bibnamefont {Hanson}}, \ and\ \bibinfo
  {author} {\bibfnamefont {A.~C.}\ \bibnamefont {Gossard}},\ }\href {\doibase
  10.1103/PhysRevLett.107.030506} {\bibfield  {journal} {\bibinfo  {journal}
  {Phys. Rev. Lett.}\ }\textbf {\bibinfo {volume} {107}},\ \bibinfo {pages}
  {030506} (\bibinfo {year} {2011})}\BibitemShut {NoStop}%
\bibitem [{\citenamefont {Gilbert}\ \emph {et~al.}(2002)\citenamefont
  {Gilbert}, \citenamefont {Akis},\ and\ \citenamefont
  {Ferry}}]{ApplPhysLett.81.22}%
  \BibitemOpen
  \bibfield  {author} {\bibinfo {author} {\bibfnamefont {M.~J.}\ \bibnamefont
  {Gilbert}}, \bibinfo {author} {\bibfnamefont {R.}~\bibnamefont {Akis}}, \
  and\ \bibinfo {author} {\bibfnamefont {D.~K.}\ \bibnamefont {Ferry}},\
  }\href@noop {} {\bibfield  {journal} {\bibinfo  {journal} {Appl. Phys.
  Lett.}\ }\textbf {\bibinfo {volume} {81}},\ \bibinfo {pages} {4284} (\bibinfo
  {year} {2002})}\BibitemShut {NoStop}%
\bibitem [{\citenamefont {Ionicioiu}\ \emph {et~al.}(2001)\citenamefont
  {Ionicioiu}, \citenamefont {Amaratunga},\ and\ \citenamefont
  {Udrea}}]{Ionicioiu.15.125}%
  \BibitemOpen
  \bibfield  {author} {\bibinfo {author} {\bibfnamefont {R.}~\bibnamefont
  {Ionicioiu}}, \bibinfo {author} {\bibfnamefont {G.}~\bibnamefont
  {Amaratunga}}, \ and\ \bibinfo {author} {\bibfnamefont {F.}~\bibnamefont
  {Udrea}},\ }\href@noop {} {\bibfield  {journal} {\bibinfo  {journal} {Int. J.
  of Mod. Phys. B}\ }\textbf {\bibinfo {volume} {15}},\ \bibinfo {pages} {125}
  (\bibinfo {year} {2001})}\BibitemShut {NoStop}%
\bibitem [{\citenamefont {Bertoni}\ \emph {et~al.}(2000)\citenamefont
  {Bertoni}, \citenamefont {Bordone}, \citenamefont {Brunetti}, \citenamefont
  {Jacoboni},\ and\ \citenamefont {Reggiani}}]{PhysRevLett.84.5912}%
  \BibitemOpen
  \bibfield  {author} {\bibinfo {author} {\bibfnamefont {A.}~\bibnamefont
  {Bertoni}}, \bibinfo {author} {\bibfnamefont {P.}~\bibnamefont {Bordone}},
  \bibinfo {author} {\bibfnamefont {R.}~\bibnamefont {Brunetti}}, \bibinfo
  {author} {\bibfnamefont {C.}~\bibnamefont {Jacoboni}}, \ and\ \bibinfo
  {author} {\bibfnamefont {S.}~\bibnamefont {Reggiani}},\ }\href {\doibase
  10.1103/PhysRevLett.84.5912} {\bibfield  {journal} {\bibinfo  {journal}
  {Phys. Rev. Lett.}\ }\textbf {\bibinfo {volume} {84}},\ \bibinfo {pages}
  {5912} (\bibinfo {year} {2000})}\BibitemShut {NoStop}%
\bibitem [{\citenamefont {Ferry}\ \emph {et~al.}(2001)\citenamefont {Ferry},
  \citenamefont {Akis},\ and\ \citenamefont
  {Harris}}]{SpperlatticeandMicrostructure.2.30}%
  \BibitemOpen
  \bibfield  {author} {\bibinfo {author} {\bibfnamefont {D.~K.}\ \bibnamefont
  {Ferry}}, \bibinfo {author} {\bibfnamefont {R.}~\bibnamefont {Akis}}, \ and\
  \bibinfo {author} {\bibfnamefont {J.}~\bibnamefont {Harris}},\ }\href@noop {}
  {\bibfield  {journal} {\bibinfo  {journal} {Spperlattice and Microstructure}\
  }\textbf {\bibinfo {volume} {20}} (\bibinfo {year} {2001})}\BibitemShut
  {NoStop}%
\bibitem [{\citenamefont {Marchi}\ \emph {et~al.}(2003)\citenamefont {Marchi},
  \citenamefont {Bertoni}, \citenamefont {Reggiani},\ and\ \citenamefont
  {Rudan}}]{NANOTECHNOLOGYIEEE.1.3}%
  \BibitemOpen
  \bibfield  {author} {\bibinfo {author} {\bibfnamefont {A.}~\bibnamefont
  {Marchi}}, \bibinfo {author} {\bibfnamefont {A.}~\bibnamefont {Bertoni}},
  \bibinfo {author} {\bibfnamefont {S.}~\bibnamefont {Reggiani}}, \ and\
  \bibinfo {author} {\bibfnamefont {M.}~\bibnamefont {Rudan}},\ }\href@noop {}
  {\bibfield  {journal} {\bibinfo  {journal} {IEEE Transaction on
  nanotechnology}\ }\textbf {\bibinfo {volume} {3}},\ \bibinfo {pages} {129}
  (\bibinfo {year} {2003})}\BibitemShut {NoStop}%
\bibitem [{\citenamefont {Snyder}\ and\ \citenamefont
  {Reichl}(2004)}]{PhysRevA.70.052330}%
  \BibitemOpen
  \bibfield  {author} {\bibinfo {author} {\bibfnamefont {M.~G.}\ \bibnamefont
  {Snyder}}\ and\ \bibinfo {author} {\bibfnamefont {L.~E.}\ \bibnamefont
  {Reichl}},\ }\href {\doibase 10.1103/PhysRevA.70.052330} {\bibfield
  {journal} {\bibinfo  {journal} {Phys. Rev. A}\ }\textbf {\bibinfo {volume}
  {70}},\ \bibinfo {pages} {052330} (\bibinfo {year} {2004})}\BibitemShut
  {NoStop}%
\bibitem [{\citenamefont {Harris}\ \emph {et~al.}(2001)\citenamefont {Harris},
  \citenamefont {Akis},\ and\ \citenamefont {Ferry}}]{ApplPhysLett.79.14}%
  \BibitemOpen
  \bibfield  {author} {\bibinfo {author} {\bibfnamefont {J.}~\bibnamefont
  {Harris}}, \bibinfo {author} {\bibfnamefont {R.}~\bibnamefont {Akis}}, \ and\
  \bibinfo {author} {\bibfnamefont {D.~K.}\ \bibnamefont {Ferry}},\ }\href@noop
  {} {\bibfield  {journal} {\bibinfo  {journal} {Appl. Phys. Lett.}\ }\textbf
  {\bibinfo {volume} {79}},\ \bibinfo {pages} {2214} (\bibinfo {year}
  {2001})}\BibitemShut {NoStop}%
\bibitem [{\citenamefont {Ramamoorthy}\ \emph {et~al.}(2006)\citenamefont
  {Ramamoorthy}, \citenamefont {Akis},\ and\ \citenamefont
  {Bird}}]{NANOTECHNOLOGYIEEE.6.5}%
  \BibitemOpen
  \bibfield  {author} {\bibinfo {author} {\bibfnamefont {A.}~\bibnamefont
  {Ramamoorthy}}, \bibinfo {author} {\bibfnamefont {R.}~\bibnamefont {Akis}}, \
  and\ \bibinfo {author} {\bibfnamefont {J.~P.}\ \bibnamefont {Bird}},\
  }\href@noop {} {\bibfield  {journal} {\bibinfo  {journal} {IEEE Transaction
  on nanotechnology}\ }\textbf {\bibinfo {volume} {5}},\ \bibinfo {pages} {712}
  (\bibinfo {year} {2006})}\BibitemShut {NoStop}%
\bibitem [{\citenamefont {Pingue}\ \emph {et~al.}(2005)\citenamefont {Pingue},
  \citenamefont {Piazza},\ and\ \citenamefont {F.}}]{ApplPhysLett.86.052102}%
  \BibitemOpen
  \bibfield  {author} {\bibinfo {author} {\bibfnamefont {P.}~\bibnamefont
  {Pingue}}, \bibinfo {author} {\bibfnamefont {V.}~\bibnamefont {Piazza}}, \
  and\ \bibinfo {author} {\bibfnamefont {B.}~\bibnamefont {F.}},\ }\href@noop
  {} {\bibfield  {journal} {\bibinfo  {journal} {Appl. Phys. Lett.}\ }\textbf
  {\bibinfo {volume} {86}},\ \bibinfo {pages} {052102} (\bibinfo {year}
  {2005})}\BibitemShut {NoStop}%
\bibitem [{\citenamefont {Bordone}\ \emph {et~al.}(2004)\citenamefont
  {Bordone}, \citenamefont {Bertoni}, \citenamefont {Rosini},\ and\
  \citenamefont {Jacoboni}}]{SemicondSciTechnol.19.412}%
  \BibitemOpen
  \bibfield  {author} {\bibinfo {author} {\bibfnamefont {P.}~\bibnamefont
  {Bordone}}, \bibinfo {author} {\bibfnamefont {A.}~\bibnamefont {Bertoni}},
  \bibinfo {author} {\bibfnamefont {M.}~\bibnamefont {Rosini}}, \ and\ \bibinfo
  {author} {\bibfnamefont {C.}~\bibnamefont {Jacoboni}},\ }\href@noop {}
  {\bibfield  {journal} {\bibinfo  {journal} {Semicond.Sci.Technol.}\ }\textbf
  {\bibinfo {volume} {19}},\ \bibinfo {pages} {412} (\bibinfo {year}
  {2004})}\BibitemShut {NoStop}%
\bibitem [{\citenamefont {Gudmundsson}\ \emph {et~al.}(2009)\citenamefont
  {Gudmundsson}, \citenamefont {Gainar}, \citenamefont {Tang}, \citenamefont
  {Moldoveanu},\ and\ \citenamefont {Manolecu}}]{Vidar11.113007}%
  \BibitemOpen
  \bibfield  {author} {\bibinfo {author} {\bibfnamefont {V.}~\bibnamefont
  {Gudmundsson}}, \bibinfo {author} {\bibfnamefont {C.}~\bibnamefont {Gainar}},
  \bibinfo {author} {\bibfnamefont {C.-S.}\ \bibnamefont {Tang}}, \bibinfo
  {author} {\bibfnamefont {V.}~\bibnamefont {Moldoveanu}}, \ and\ \bibinfo
  {author} {\bibfnamefont {A.}~\bibnamefont {Manolecu}},\ }\href@noop {}
  {\bibfield  {journal} {\bibinfo  {journal} {New J. Phys.}\ }\textbf {\bibinfo
  {volume} {11}},\ \bibinfo {pages} {113007} (\bibinfo {year}
  {2009})}\BibitemShut {NoStop}%
\bibitem [{\citenamefont {Abdullah}\ \emph {et~al.}(2010)\citenamefont
  {Abdullah}, \citenamefont {Tang},\ and\ \citenamefont
  {Gudmundsson}}]{PhysRevB.82.195325}%
  \BibitemOpen
  \bibfield  {author} {\bibinfo {author} {\bibfnamefont {N.~R.}\ \bibnamefont
  {Abdullah}}, \bibinfo {author} {\bibfnamefont {C.-S.}\ \bibnamefont {Tang}},
  \ and\ \bibinfo {author} {\bibfnamefont {V.}~\bibnamefont {Gudmundsson}},\
  }\href {\doibase 10.1103/PhysRevB.82.195325} {\bibfield  {journal} {\bibinfo
  {journal} {Phys. Rev. B}\ }\textbf {\bibinfo {volume} {82}},\ \bibinfo
  {pages} {195325} (\bibinfo {year} {2010})}\BibitemShut {NoStop}%
\bibitem [{\citenamefont {Gudmundsson}\ \emph {et~al.}(2012)\citenamefont
  {Gudmundsson}, \citenamefont {Jonasson}, \citenamefont {Tang}, \citenamefont
  {Goan},\ and\ \citenamefont {Manolescu}}]{Vidar85.075306}%
  \BibitemOpen
  \bibfield  {author} {\bibinfo {author} {\bibfnamefont {V.}~\bibnamefont
  {Gudmundsson}}, \bibinfo {author} {\bibfnamefont {O.}~\bibnamefont
  {Jonasson}}, \bibinfo {author} {\bibfnamefont {C.-S.}\ \bibnamefont {Tang}},
  \bibinfo {author} {\bibfnamefont {H.-S.}\ \bibnamefont {Goan}}, \ and\
  \bibinfo {author} {\bibfnamefont {A.}~\bibnamefont {Manolescu}},\ }\href
  {\doibase 10.1103/PhysRevB.85.075306} {\bibfield  {journal} {\bibinfo
  {journal} {Phys. Rev. B}\ }\textbf {\bibinfo {volume} {85}},\ \bibinfo
  {pages} {075306} (\bibinfo {year} {2012})}\BibitemShut {NoStop}%
\bibitem [{\citenamefont {Jaynes}\ and\ \citenamefont
  {Cummings}(1963)}]{ProceedingsoftheIEEE.51.89}%
  \BibitemOpen
  \bibfield  {author} {\bibinfo {author} {\bibfnamefont {E.~T.}\ \bibnamefont
  {Jaynes}}\ and\ \bibinfo {author} {\bibfnamefont {F.~W.}\ \bibnamefont
  {Cummings}},\ }\href@noop {} {\bibfield  {journal} {\bibinfo  {journal}
  {Proceedings of the IEEE}\ }\textbf {\bibinfo {volume} {51}},\ \bibinfo
  {pages} {89} (\bibinfo {year} {1963})}\BibitemShut {NoStop}%
\bibitem [{\citenamefont {Yannouleas}\ and\ \citenamefont
  {Landman}(2007)}]{Yannouleas70.2067}%
  \BibitemOpen
  \bibfield  {author} {\bibinfo {author} {\bibfnamefont {C.}~\bibnamefont
  {Yannouleas}}\ and\ \bibinfo {author} {\bibfnamefont {U.}~\bibnamefont
  {Landman}},\ }\href@noop {} {\bibfield  {journal} {\bibinfo  {journal} {Rep.
  Prog. Phys}\ }\textbf {\bibinfo {volume} {70}},\ \bibinfo {pages} {2067}
  (\bibinfo {year} {2007})}\BibitemShut {NoStop}%
\bibitem [{\citenamefont {Jonasson}\ \emph {et~al.}(2012)\citenamefont
  {Jonasson}, \citenamefont {Tang}, \citenamefont {Goan}, \citenamefont
  {Manolescu},\ and\ \citenamefont
  {Gudmundsson}}]{NewJournalOfPhysics.14.013036}%
  \BibitemOpen
  \bibfield  {author} {\bibinfo {author} {\bibfnamefont {O.}~\bibnamefont
  {Jonasson}}, \bibinfo {author} {\bibfnamefont {C.-S.}\ \bibnamefont {Tang}},
  \bibinfo {author} {\bibfnamefont {H.-S.}\ \bibnamefont {Goan}}, \bibinfo
  {author} {\bibfnamefont {A.}~\bibnamefont {Manolescu}}, \ and\ \bibinfo
  {author} {\bibfnamefont {V.}~\bibnamefont {Gudmundsson}},\ }\href@noop {}
  {\bibfield  {journal} {\bibinfo  {journal} {New journal of physics}\ }\textbf
  {\bibinfo {volume} {14}},\ \bibinfo {pages} {013036} (\bibinfo {year}
  {2012})}\BibitemShut {NoStop}%
\bibitem [{\citenamefont {Gudmundsson}\ \emph {et~al.}(2013)\citenamefont
  {Gudmundsson}, \citenamefont {Jonasson}, \citenamefont {Arnold},
  \citenamefont {Tang}, \citenamefont {Goan},\ and\ \citenamefont
  {Manolescu}}]{Vidar61.305}%
  \BibitemOpen
  \bibfield  {author} {\bibinfo {author} {\bibfnamefont {V.}~\bibnamefont
  {Gudmundsson}}, \bibinfo {author} {\bibfnamefont {O.}~\bibnamefont
  {Jonasson}}, \bibinfo {author} {\bibfnamefont {T.}~\bibnamefont {Arnold}},
  \bibinfo {author} {\bibfnamefont {C.-S.}\ \bibnamefont {Tang}}, \bibinfo
  {author} {\bibfnamefont {H.-S.}\ \bibnamefont {Goan}}, \ and\ \bibinfo
  {author} {\bibfnamefont {A.}~\bibnamefont {Manolescu}},\ }\href@noop {}
  {\bibfield  {journal} {\bibinfo  {journal} {Fortschr. Phys.}\ }\textbf
  {\bibinfo {volume} {61}},\ \bibinfo {pages} {305} (\bibinfo {year}
  {2013})}\BibitemShut {NoStop}%
\bibitem [{\citenamefont {Sornborger}\ \emph {et~al.}(2004)\citenamefont
  {Sornborger}, \citenamefont {Cleland},\ and\ \citenamefont
  {Geller}}]{PhysRevA.70.052315}%
  \BibitemOpen
  \bibfield  {author} {\bibinfo {author} {\bibfnamefont {A.~T.}\ \bibnamefont
  {Sornborger}}, \bibinfo {author} {\bibfnamefont {A.~N.}\ \bibnamefont
  {Cleland}}, \ and\ \bibinfo {author} {\bibfnamefont {M.~R.}\ \bibnamefont
  {Geller}},\ }\href {\doibase 10.1103/PhysRevA.70.052315} {\bibfield
  {journal} {\bibinfo  {journal} {Phys. Rev. A}\ }\textbf {\bibinfo {volume}
  {70}},\ \bibinfo {pages} {052315} (\bibinfo {year} {2004})}\BibitemShut
  {NoStop}%
\bibitem [{\citenamefont {Wu}\ and\ \citenamefont
  {Yang}(2007)}]{PhysRevLett.98.013601}%
  \BibitemOpen
  \bibfield  {author} {\bibinfo {author} {\bibfnamefont {Y.}~\bibnamefont
  {Wu}}\ and\ \bibinfo {author} {\bibfnamefont {X.}~\bibnamefont {Yang}},\
  }\href {\doibase 10.1103/PhysRevLett.98.013601} {\bibfield  {journal}
  {\bibinfo  {journal} {Phys. Rev. Lett.}\ }\textbf {\bibinfo {volume} {98}},\
  \bibinfo {pages} {013601} (\bibinfo {year} {2007})}\BibitemShut {NoStop}%
\bibitem [{\citenamefont {Breuer}\ and\ \citenamefont
  {Petruccione}(2002)}]{Breuer2002}%
  \BibitemOpen
  \bibfield  {author} {\bibinfo {author} {\bibfnamefont {H.-P.}\ \bibnamefont
  {Breuer}}\ and\ \bibinfo {author} {\bibfnamefont {F.}~\bibnamefont
  {Petruccione}},\ }\href@noop {} {\bibfield  {journal} {\bibinfo  {journal}
  {\textit{The Theory of Open Quantum Systems}}\ } (\bibinfo {year} {Oxford
  University Press, Oxford, 2002})}\BibitemShut {NoStop}%
\bibitem [{\citenamefont {Esposito}\ \emph {et~al.}(2009)\citenamefont
  {Esposito}, \citenamefont {Harbola},\ and\ \citenamefont
  {Mukamel}}]{RevModPhys.81.1665}%
  \BibitemOpen
  \bibfield  {author} {\bibinfo {author} {\bibfnamefont {M.}~\bibnamefont
  {Esposito}}, \bibinfo {author} {\bibfnamefont {U.}~\bibnamefont {Harbola}}, \
  and\ \bibinfo {author} {\bibfnamefont {S.}~\bibnamefont {Mukamel}},\ }\href
  {\doibase 10.1103/RevModPhys.81.1665} {\bibfield  {journal} {\bibinfo
  {journal} {Rev. Mod. Phys.}\ }\textbf {\bibinfo {volume} {81}},\ \bibinfo
  {pages} {1665} (\bibinfo {year} {2009})}\BibitemShut {NoStop}%
\bibitem [{\citenamefont {Abdullah}\ \emph {et~al.}(2013)\citenamefont
  {Abdullah}, \citenamefont {Tang}, \citenamefont {Manolescu},\ and\
  \citenamefont {Gudmundsson}}]{Nzar.25.465302}%
  \BibitemOpen
  \bibfield  {author} {\bibinfo {author} {\bibfnamefont {N.~R.}\ \bibnamefont
  {Abdullah}}, \bibinfo {author} {\bibfnamefont {C.~S.}\ \bibnamefont {Tang}},
  \bibinfo {author} {\bibfnamefont {A.}~\bibnamefont {Manolescu}}, \ and\
  \bibinfo {author} {\bibfnamefont {V.}~\bibnamefont {Gudmundsson}},\
  }\href@noop {} {\bibfield  {journal} {\bibinfo  {journal} {Journal of
  Physics:Condensed Matter}\ }\textbf {\bibinfo {volume} {25}},\ \bibinfo
  {pages} {465302} (\bibinfo {year} {2013})}\BibitemShut {NoStop}%
\bibitem [{\citenamefont {Haake}(p 98)}]{Haake1973}%
  \BibitemOpen
  \bibfield  {author} {\bibinfo {author} {\bibfnamefont {F.}~\bibnamefont
  {Haake}},\ }\href@noop {} {\bibfield  {journal} {\bibinfo  {journal}
  {\textit{Quantum Statistics in Optics and Solid-state Physics}, edited by G.
  Hohler and E.A. Niekisch, Springer Tracts in Modern Physics Vol.}\ }\textbf
  {\bibinfo {volume} {66}} (\bibinfo {year} {Springer, Berlin, Heidelberg, New
  York, 1973, p. 98.})}\BibitemShut {NoStop}%
\bibitem [{\citenamefont {Moldovean}\ \emph {et~al.}(2009)\citenamefont
  {Moldovean}, \citenamefont {Manolescu},\ and\ \citenamefont
  {Gudmundsson}}]{NewJournalOfPhysics.11.073019}%
  \BibitemOpen
  \bibfield  {author} {\bibinfo {author} {\bibfnamefont {V.}~\bibnamefont
  {Moldovean}}, \bibinfo {author} {\bibfnamefont {A.}~\bibnamefont
  {Manolescu}}, \ and\ \bibinfo {author} {\bibfnamefont {V.}~\bibnamefont
  {Gudmundsson}},\ }\href@noop {} {\bibfield  {journal} {\bibinfo  {journal}
  {New journal of physics}\ }\textbf {\bibinfo {volume} {11}},\ \bibinfo
  {pages} {073019} (\bibinfo {year} {2009})}\BibitemShut {NoStop}%
\bibitem [{\citenamefont {Datta}()}]{Datta1995}%
  \BibitemOpen
  \bibfield  {author} {\bibinfo {author} {\bibfnamefont {S.}~\bibnamefont
  {Datta}},\ }\href@noop {} {\bibinfo  {journal} {Electronic Transport in
  Mesoscopic system (Cambridge University Press, Cambridge, 1995)}\
  }\BibitemShut {NoStop}%
\bibitem [{\citenamefont {Abdullah}\ \emph
  {et~al.}(2014{\natexlab{a}})\citenamefont {Abdullah}, \citenamefont {Tang},
  \citenamefont {Manolescu},\ and\ \citenamefont
  {Gudmundsson}}]{arXiv:1408.1007}%
  \BibitemOpen
\bibfield  {journal} {  }\bibfield  {author} {\bibinfo {author} {\bibfnamefont
  {N.~R.}\ \bibnamefont {Abdullah}}, \bibinfo {author} {\bibfnamefont {C.~S.}\
  \bibnamefont {Tang}}, \bibinfo {author} {\bibfnamefont {A.}~\bibnamefont
  {Manolescu}}, \ and\ \bibinfo {author} {\bibfnamefont {V.}~\bibnamefont
  {Gudmundsson}},\ }\href@noop {} {\bibfield  {journal} {\bibinfo  {journal}
  {arXiv:1408.1007}\ } (\bibinfo {year} {2014}{\natexlab{a}})}\BibitemShut
  {NoStop}%
\bibitem [{\citenamefont {Zibold}\ \emph {et~al.}(2007)\citenamefont {Zibold},
  \citenamefont {Vogl},\ and\ \citenamefont {Bertoni}}]{PhysRevB.76.195301}%
  \BibitemOpen
  \bibfield  {author} {\bibinfo {author} {\bibfnamefont {T.}~\bibnamefont
  {Zibold}}, \bibinfo {author} {\bibfnamefont {P.}~\bibnamefont {Vogl}}, \ and\
  \bibinfo {author} {\bibfnamefont {A.}~\bibnamefont {Bertoni}},\ }\href
  {\doibase 10.1103/PhysRevB.76.195301} {\bibfield  {journal} {\bibinfo
  {journal} {Phys. Rev. B}\ }\textbf {\bibinfo {volume} {76}},\ \bibinfo
  {pages} {195301} (\bibinfo {year} {2007})}\BibitemShut {NoStop}%
\bibitem [{\citenamefont {Gong}\ \emph {et~al.}(2007)\citenamefont {Gong},
  \citenamefont {Yang},\ and\ \citenamefont {Feng}}]{ChinPhysLett.24.8}%
  \BibitemOpen
  \bibfield  {author} {\bibinfo {author} {\bibfnamefont {J.}~\bibnamefont
  {Gong}}, \bibinfo {author} {\bibfnamefont {F.-H.}\ \bibnamefont {Yang}}, \
  and\ \bibinfo {author} {\bibfnamefont {S.-L.}\ \bibnamefont {Feng}},\
  }\href@noop {} {\bibfield  {journal} {\bibinfo  {journal} {Chin. Phys.
  Lett.}\ }\textbf {\bibinfo {volume} {24}},\ \bibinfo {pages} {2383} (\bibinfo
  {year} {2007})}\BibitemShut {NoStop}%
\bibitem [{\citenamefont {Abdullah}\ \emph
  {et~al.}(2014{\natexlab{b}})\citenamefont {Abdullah}, \citenamefont {Tang},
  \citenamefont {Manolescu},\ and\ \citenamefont
  {Gudmundsson}}]{PhysicaE.64.254}%
  \BibitemOpen
  \bibfield  {author} {\bibinfo {author} {\bibfnamefont {N.~R.}\ \bibnamefont
  {Abdullah}}, \bibinfo {author} {\bibfnamefont {C.~S.}\ \bibnamefont {Tang}},
  \bibinfo {author} {\bibfnamefont {A.}~\bibnamefont {Manolescu}}, \ and\
  \bibinfo {author} {\bibfnamefont {V.}~\bibnamefont {Gudmundsson}},\
  }\href@noop {} {\bibfield  {journal} {\bibinfo  {journal} {Physica E}\
  }\textbf {\bibinfo {volume} {64}},\ \bibinfo {pages} {254} (\bibinfo {year}
  {2014}{\natexlab{b}})}\BibitemShut {NoStop}%
\end{thebibliography}

%
%
\end{document}